\documentstyle[12pt,equation]{article} 
 1                                        
\begin{document}                                                              
\begin{center}                                                                
{ \large\bf THE ELECTROMAGNETIC COUPLING IN KEMMER-DUFFIN-PETIAU THEORY \\}
\vskip 2cm  
{ Marek Nowakowski \\}                              
Grup de F\'{i}sica 
Te\`orica, Universitat Aut\`onoma de Barcelona, 08193 Bellaterra,
Spain 
\end{center}
\vskip .5cm                             
\begin{abstract}
We analyse the electromagnetic coupling in the Kemmer-Duffin-Petiau
(KDP) equation. Since the KDP--equation which describes spin-$0$
and spin-$1$ bosons is of Dirac-type,
we examine some analogies and differences from the Dirac equation. 
The main difference to the Dirac equation is that the KDP equation contains 
redundant components. 
We will show that as a result certain interaction terms in
the Hamilton form of the KDP equation do not have a physical meaning and will
not affect the calculation of physical observables. We point
out that a second order KDP equation derived by Kemmer as an analogy to the
second order Dirac equation is of limited physical applicability as (i)
it belongs to a class of second order equations which can be derived from the
original KDP equation and (ii) it lacks a back--transformation which would
allow one to obtain solutions of the KDP equation out of solutions of 
the second order equation. We therefore 
suggest a different higher order equation which, as far as the
solutions for the wave functions are concerned, is equivalent to the orginal
first order KDP wave equation.
\end{abstract}                                                                
\newpage 
\section {Introduction}

Since the early days of Quantum Mechanics physicists have tried to construct
relativistically invariant wave equations for different spins and in different
disguises. Beside the, by now standard, Klein-Gordon and Dirac equations
there are other Lorentz invariant wave equations. We mention here the 
infinite component equations of Majorana type which describe a tower of higher
spin states \cite{majorana}, an equation suggested by Dirac for spin-0
bosons \cite{dirac} and its generalizations \cite{mukunda},
the six dimensional Weinberg--Shay--Good equation for spin-1 bosons 
\cite{weinberg} and the Kemmer equation, also called 
Kemmer-Duffin-Petiau equation \cite{kemmer}, for spin-0 and spin-1 bosons.
It is the last set of  
equations which, in the presence of an electromagnetc
field, we will investigate here more closely. 
Although we have a consistent desription of bosons in quantum field theory in
terms of the Klein--Gordon equation it is 
often instructive and useful to look at
the same problem form a yet different point of view, 
in the  case of concern here from the perspective 
of the KDP equation.     
Indeed, as we will sketch below, new insights can still be gained.
The KDP equation which has been known for some time is an equation of
Dirac type
\begin{equation} \label{0}
(i\beta_{\mu}\partial^{\mu}+m)\psi=0
\end{equation}  
where the matrices $\beta_{\mu}$ satisfy a trilinear algebra  \cite{duffin}
\begin{equation} \label{1}
\beta^{\mu}\beta^{\nu}\beta^{\alpha}+\beta^{\alpha}\beta^{\nu}\beta^{\mu}
=g^{\mu \nu}\beta^{\alpha}+g^{\nu \alpha}\beta^{\mu}
\end{equation}
The convention for the metric tensor is here $g^{\mu \nu}=diag(1,-1,-1,-1)$.
The algebra (1.2) has three disctinct representions: the trivial one with
$\beta_{\mu}=0$, a five-dimensional representation describing spin-0 particles
and a ten dimensional one for the spin-1 bosons. One can derive (1.1) from a
Lagrangian, construct interaction 
Lagrangians with other fields and quantize the
theory. An equation similar to (1.1) can be also written down for the massless
case. We will not dwell here on these issues and refer the interested reader
to \cite{chandra} and \cite{umezawa} where these topics are treated in more
details.

The renewed interest in (1.1) can be attributed to the discovery of a new
conserved four-vector current whose zeroth component is positive definite
and can therefore
be interpreted as a pobability density to find a particle at a point
$r$ \cite{ghose1}, \cite{ghose2}, \cite{ghose3}. 
Strictly speaking from (1.1) one can construct two conserved
four-vector currents. The first one
\begin{eqnarray} \label{2}
j^{\mu} \equiv \bar{\psi}\beta^{\mu}\psi \nonumber \\
\partial_{\mu}j^{\mu}=0
\end{eqnarray}
with $\bar{\psi}=\psi^{\dagger}(2\beta_0^2-1)$ shares with Klein-Gordon theory
the property that $j_0$ is not positive definite. Defining 
\begin{equation} \label{3}
\phi \equiv {\psi \over \sqrt{\int d^3x \psi^{\dagger}\psi}}
\end{equation}
one can also derive from (1.1) a different conservation law in the form
\begin{equation} \label{4}
\partial_{\mu}s^{\mu} \equiv \partial_0(\phi^{\dagger}\phi) 
+ \partial_i(\phi^{\dagger}
\tilde{\beta}_i \phi)=0
\end{equation}
where $\tilde{\beta}_i=[\beta_0, \beta_i]$. Based on an idea of Holland 
\cite{holland} to use
the symmetric
energy-momentum tensor $\theta_{\mu \nu}$ to construct conserved four-vectors
Ghose, Home and Roy were able to give a proof 
that $s^{\mu}=(s^0, s^i)$ indeed transforms as a four vector 
\cite{ghose1,ghose2, ghose3} . 
In contrast
to (1.3) we have now clearly $s^0 >0$. It would be probably tedious, if not
impossible, to establish such a result in the Klein-Gordon theory. 
In this letter we will not examine
further consequences of the current (1.5), 
but would only like to point out that
such a new current is welcome and useful especially for neutral particles. For
instance, in \cite{us}
a non-relativistic current for neutral kaons has been used to
compute space (in contrast to time) dependent conversion probabilities
$P_{K^0 \to \bar{K^0}}(r)$ and $P_{K^0 \to K^0}(r)$ by integrating the current
over time and surface through which the particles pass. 
With the help of (1.5) the same exercise could be
now done for a general relativistic current \cite{me}. 
This is not a mere academic
problem as the ongoing discussion about the interference phase appearing
in the conversion probabilities shows (for details see \cite{us}).

Another interesting aspect which will be the main subject of the present
paper is the external field problem in connection with (1.1). An
external field problem in the context of a
relativistic equation is only well defined provided the
external field is not too strong. But once we restrict ourselves 
to weak fields, 
useful results for energy eigenvalues and for wave functions can be obtained.
This is the case with the Dirac equation in a homogeneous magnetic field and
a plane wave electromagnetic field (see e.g. \cite{landau}). 

The interest in the external field problem in the KDP equation is twofold.
First, since (1.1) looks formally like the Dirac equation, it is useful
for practical purposes (like methods to solve 
the KDP equation, etc) to see how far
one can stretch the analogy to the Dirac equation. Note that because of the
existence of the current (1.5) a relativistic quantum mechanics of bosons
is now possible.
Secondly and more
importantly, 
it is not a priori obvious if the KDP equation including interactions
is equivalent to the Klein-Gordon description (it is easy to show that in the
free case both theories are equivalent)\footnote{It might be also worthwhile
to remark that in view of (1.1) certain mixing phenomena, like $\omega^0-
\rho^0$ and $\eta - \eta'$ mixing, when formulated in the Lagrangain language,
could be now linear, as opposed to quadratic, in the mass parameters}.
When we employ the minimal coupling
scheme for the electromagnetic four--potential $A_{\mu}$ in eq. (1.1) and
go over to the Hamilton formulation of the theory we will discover
a term which seems to lack a physical interpretation. Kemmer has discussed
a similar looking term in a second order Klein-Gordon type of equation which
in analogy to second order Dirac equation (see \cite{landau}) he derived from 
(1.1).  This has led to a suggestion to couple the electromagnetic
field only at the level of the Hamilton form which would then avoid the
cumbersome term \cite{ghose3}. 
It will be shown, however, that this cumbersome term has no
physical meaning and it vanishes when the first order 
wave equation is reduced to its
physical components. Moreover, the analogy of the second order Kemmer equation
to the similar looking second order Dirac equation is very limited. Whereas in
the latter case we can transform solutions of the second order equation to
solutions of the Dirac equation and vice versa, such a one-to-one correspondence
is not possible in the Kemmer case. More seriously the second order Kemmer
equation is only one member of 
a class of second order equations which, in principle, can be derived
from (1.1). Their physical significance is therefore not clear.
On account of this we suggest a higher order wave equation which has the same
virtue as the second order Dirac equation (i.e. a back-transformation to the
solutions of the first order equation), but does not, in the free case, reduce
to the Klein-Gordon equation. This in turn implies that we should not expect
that every term in such an equation has an interpretable meaning.

\setcounter{equation}{0}
\section{Algebra}

Before discussing the external field problem it is convenient for calculational
as well as physical reasons to separate the spin-0 case from the spin-1 case
by purely algebraic means. In \cite{chandra} 
it was shown that given the algebra ({1.2)
there exists a matrix $\beta$ satisfying
\begin{eqnarray}\label{5}
\beta^2&=&\beta \nonumber \\
\{\beta , \beta_{\mu}\}&=&\beta_{\mu}
\end{eqnarray}
Obviously $\beta'=1-\beta$ will also satisfy (2.1). For the spin-0 case it
is then possible to give a relation which is stronger than (1.2), namely
\cite{chandra}
\begin{equation} \label{ex1}
\beta^{\mu}\beta^{\nu}\beta^{\alpha}=g^{\mu \nu}\beta^{\alpha}\beta
+g^{\nu \alpha}\beta \beta^{\mu}
\end{equation}
It is easy to show that (1.2) follows from (2.2). We supplement
this remark by observing that the spin-1 algebra can be also characterized
by a algebraic relation different from (1.2). Defining the matrix $\omega$
by 
\begin{equation} \label{6}
\omega \equiv {i \over 4} \epsilon^{\mu \nu \sigma \delta}
\beta_{\mu}\beta_{\nu}\beta_{\sigma}\beta_{\delta}
\end{equation}
($\epsilon^{0123}=1$) we see that $\omega=0$ for spin-0 on account of
(2.2). One can deduce then several properties of $\omega$ \cite{chandra}
out of which we quote
\begin{eqnarray} \label{7}
\omega^2=1-\beta \\
\{\omega^2 , \beta_{\mu}\}=\beta_{\mu} \\
\beta_{\mu}\omega \beta_{\nu} + \beta_{\nu}\omega \beta_{\mu}=0 \\
\beta_{\mu}\beta_{\nu}\omega + \omega\beta_{\nu}\beta_{\mu}=g_{\mu \nu}
\omega
\end{eqnarray}
The matrix $\omega$ has a close analogy to the $\gamma_5$ matrix in the 
Clifford algebra and can be used to extend (1.2) to five dimensional
space-time.
Using now (2.5)-(2.7) we get
\begin{eqnarray} \label{8}
\beta^{\nu}\beta^{\alpha}\beta^{\mu}&=& g^{\alpha \mu}\beta^{\nu}\omega^2
+ g^{\nu \alpha}\omega^2 \beta^{\mu} \nonumber \\
&+& \beta^{\mu}\omega \beta^{\nu}\beta^{\alpha}\omega + \omega \beta^{\alpha}
\beta^{\mu}\omega \beta^{\nu}
\end{eqnarray}
With the help of eq. (2.7) it is 
not difficult to see that (2.8) leads again to the original algebra (1.2).
Hence we can state that 
the ($\beta$,$\beta_{\mu}$) algebra (2.2) characterizes the
matrices for the spin-0 case whereas
the ($\omega$,$\beta_{\mu}$) algebra does the
same for the spin-1 case.
\setcounter{equation}{0}
\section{The wave equations in an external field}

In the following we will use the minimal electromagnetic coupling scheme
with a free electromagnetic field, $\partial_{\mu}F^{\mu \nu}=0$. The covariant
derivative and the corresponding operator are then defined as usual by
\begin{eqnarray} \label{9}
{\cal D}^{\mu} \equiv i\partial^{\mu}-eA^{\mu} \nonumber \\
\Lambda ({\cal D}) \equiv \beta_{\mu}{\cal D}^{\mu} +m
\end{eqnarray}
so that the wave equation in the presence of an
 electromagnetic field reads now
\begin{equation} \label{10}
\Lambda ({\cal D})\psi =0
\end{equation}
Below we will give a brief derivation of three equations which are important
in the KDP theory. For more details we refer the reader to \cite{kemmer}.
These three equations will be the basis for our discussion of the external
field problem in the KDP theory.

By multiplying (3.2) with ${\cal D}^{\rho}\beta_{\rho}\beta_{\nu}$ we obtain
\begin{equation} \label{11}
{\cal D}_{\nu}\psi=\beta_{\rho}\beta_{\nu}{\cal D}^{\rho}\psi
+{ie \over 2m}F^{\mu \rho}\left(\beta_{\rho}\beta_{\nu}\beta_{\mu}
+\beta_{\rho}g_{\mu \nu}\right)\psi
\end{equation}
where the standard result 
$[{\cal D}^{\mu}, {\cal D}^{\rho}]=ieF^{\rho \mu}$ has been used.
Combining (3.2) with (3.3) one can cast the KDP equation into a Hamilton form
\begin{eqnarray} \label{12}
&i&\partial_0\psi=H\psi \nonumber \\
&H& ={\cal D}^i[\beta_i, \beta_0] -\beta_0m
+eA_0 +{ie \over 2m}F^{\mu \rho}(\beta_{\rho}\beta_0 \beta_{\mu}
+\beta_{\rho}g_{\mu 0})
\end{eqnarray}
Finally one can also derive from (3.2) and (3.3)
the second order equation mentioned in the
introduction. It reads \cite{kemmer}
\begin{equation} \label{13}
\Omega_1({\cal D})\psi \equiv
\left[{\cal D}^{\alpha}{\cal D}_{\alpha}-m^2 -{ie \over 2}F^{\nu \mu}S_{
\nu \mu} -{ie \over 2m}(\beta_{\rho}\beta_{\nu}\beta_{\mu}+\beta_{\rho}
g_{\mu \nu}){\cal D}^{\nu}F^{\mu \rho}\right]\psi=0
\end{equation}
where
\begin{equation} \label{14}
S_{\nu \mu}=\beta_{\nu}\beta_{\mu}-\beta_{\mu}\beta_{\nu}
\end{equation}
whose spatial components, $S_{ij}$, are related to the spin operators.
A few comments are in order now.
\begin{itemize}
\item[(i)]
The troublesome term referred to earlier appears in all three equations,
(3.3), (3.4) and (3.5) in different disguises. 
It is proportional to $ie/2m$ and is
always connected with the
tensor matrix
$$(\beta_{\rho}\beta_{\nu}\beta_{\mu}+\beta_{\rho}g_{\mu \nu})$$
Kemmer \cite{kemmer} compared (3.5) with the second order Dirac equation
(see section 5)
and pointed out that a term corresponding to $F^{\mu \nu}S_{\mu \nu}$ appears
also in the latter equation, but not the other lengthy term involving 
one and three
$\beta_{\mu}$ matrices. Equation (3.5) consists of the Klein-Gordon operator
plus terms which one could try to interprete as spin-field interaction 
\cite{kemmer} (this is certainly true for
$F^{\mu \nu}S_{\mu \nu}$, see, however, the points (ii) and (iii) below for the
rest of the terms). It is then tempting
to view (3.5) as a close analogue to the second order Dirac equation. We will
address this question in section 5 in more detail and show that this is not the
case.

\item[(ii)]
Note that in (3.5) the derivative acts on the wave function as well as on the
electromagnetic stress tensor $F^{\mu \nu}$. However, eq. (2.2)
tells us that
for the spin-0 case we have
$$(\beta_{\rho}\beta_{\nu}\beta_{\mu}+\beta_{\rho}g_{\mu \nu})
(\partial^{\nu}F^{\mu \rho})=0$$
This is not the case for spin-1 which immediately gives rise to the
question about the meaning of the derivatives of the electric and magnetic
field in eq. (3.5). Such derivatives acting on the physical electromagnetic
fields do of course enter the Maxwell equations. However, we are faced here
with the external field problem where the electromagnetic field configurations
are supposed to be given. 
\item[(iii)]
If we look at the photon interaction with bosons from the point of view of
Feynman diagrams then, already at tree level, there will be two contributions:
a one-photon exchange with derivative couplings and a two-photon contact
diagram at the level $e^2$. We would therefore expect in (3.5) a spin-field 
interaction term at the order $e^2$. Indeed a part of the terms in (3.5)
can be rewritten in such a way as to confirm this expectation. 
Using once more eq. (3.3)
we can write 
\begin{eqnarray} \label{15}
{ie \over 2m}(\beta_{\rho}\beta_{\nu}\beta_{\mu}&-&\beta_{\rho}g_{\mu \nu})
F^{\mu \rho}{\cal D}^{\nu}\psi \nonumber \\
= {e^2 \over 4m^2}F^{\alpha \gamma}F^{\mu \rho}(\beta_{\rho}\beta_{\gamma}
\beta_{\mu}\beta_{\alpha}&+&\beta_{\rho}\beta_{\gamma}g_{\mu \alpha})\psi
\nonumber \\
=-{e^2 \over 16m^2}\biggl [F^{\alpha \gamma}F^{\mu \rho}S_{\rho \mu}
S_{\alpha \gamma}&-&2 F^{\alpha \gamma}F^{\mu \rho}S_{\rho \gamma}
S_{\mu \alpha}\biggr ]\psi +{e^2 \over 2m^2}F^{\alpha \gamma}
F_{\alpha}^{{ }\rho}\beta_{\rho}\beta_{\gamma}\psi
\end{eqnarray}
It is a matter of the relative sign in 
$(\beta_{\rho}\beta_{\nu}\beta_{\mu}+\beta_{\rho}g_{\mu \nu})$ and the
derivatives acting on $F^{\mu \nu}$ that we cannot attribute the whole
term 
$(\beta_{\rho}\beta_{\nu}\beta_{\mu}+\beta_{\rho}
g_{\mu \nu}){\cal D}^{\nu}F^{\mu \rho}\psi$
to spin-field interaction at the order $e^2$.
\item[(iv)]
Eq. (3.5) follows form (3.2), but not vice versa. 
In other words, any solution of (3.2)
will be also a solution of (3.5), but the opposite is not necessarily true.
To make eq. (3.5) meaningful one would need a prescription which would convert
solutions of (3.5) to solutions of the KDP equation (3.2). 
We will address this issue in more detail
in section 5.
\item[(v)]
Finally, had we first taken a free Hamilton operator and coupled to it,
via the minimal substitution, the electromagnetic field then the term in
(3.4) proportional to $ie/2m$ would be absent \cite{ghose3}. Two possibilities
can occur now. Either the Lagrangian method and the Hamiltonian one to couple
the electromagnetic field are indeed different or the term proportional
$ie/2m$ is unphysical.
\end{itemize}

Since the wave function in (3.2) contains redundant components an important
part of the KDP theory are the constraints. They are given by
\begin{equation} \label{16}
{\cal C}[\psi] \equiv \beta_i \beta_0^2{\cal D}^i \psi + m(1-\beta_0^2)\psi
=0
\end{equation}
Eq. (3.8) is obtained from (3.2) by multiplying the latter with
$(1-\beta_0^2)$ and using the identity $(1-\beta_0^2)\beta_i=\beta_i\beta_0^2$
which follows from (1.2).

At this stage it is quite straightforward to investigate more closely the
spin-0 case. The problem of spin-1 will be discussed in the next section.
An explicit represention of the $\beta_{\mu}$ matrices for the spin-0 case
reveals that (3.8) takes the form 
\begin{equation} \label{17}
{\cal D}^i \psi +m\psi_i=0, \, \,\, \, i=1,2,3
\end{equation}
We can also eliminate $\psi_4$ by
\begin{equation} \label{18}
{\cal D}^0\psi_5+im\psi_4=0
\end{equation}
which is a simple consequence of (3.2).
In the spin-0 case we arrive then at the standard result  
for the single physical component           
$\psi_5$ 
\begin{equation} \label{19}
({\cal D}^{\alpha}{\cal D}_{\alpha}-m^2)\psi_5=0
\end{equation}
Despite the fact that each of the eqs. (3.3-3.5) constitutes a complicated
system of coupled partial differential equations, 
being of
course also valid for spin-0, the end result in terms of the physical
component is quite simple. This signals that not every term in the eqs.
(3.3-3.5) has a physical relevance. 
Below we will perform a similar reduction 
to the physical components of the spin-1 wave function and obtain an equation
which determines these components.

\setcounter{equation}{0}
\section{The reduction of the wave equation}

In the massive
spin-1 case there are three physical components which in the represention we
are using \footnote{We are using here the $\beta_0, i\beta_i$ where
the $\beta_i$ can be found in \cite{kemmer}} are given by 
$(\psi_1,\psi_2, \psi_3)$. The unphysical variables $(\psi_7, \psi_8, \psi_9)$
and $\psi_{10}$ can be eliminated by using
\begin{eqnarray} \label{20}
{\cal D}^0\left(\begin{array}{lcr} 
\psi_1 \\ \psi_2 \\ \psi_3 \end{array}\right)&-&im\left( \begin{array}{lcr}
\psi_7 \\ \psi_8 \\ \psi_9 \end{array} \right)=0 \nonumber \\
\sum_{i=1}^{3}{\cal D}^i\psi_i& +&im\psi_{10}=0
\end{eqnarray}
The first 
equation in (4.1) follows directly from the wave equation (3.2).
The second one is part of the constraints (3.8).
The other elements $(\psi_4,\psi_5,\psi_6)$ are related to $(\psi_7,\psi_8,
\psi_9)$ also 
through the constraints (3.8). For our purposes it suffices, however,
to reduce the eq. (3.2) or equivalently (3.4) to a $6 \times 6$ equation
determining the wave function $(\psi_i, \psi_j)$, i=1,2,3  and j=7,8,9.
This is exactly the wave function given by $\beta_0^2\psi$. The reduction
process of eq. (3.2) or alternatively (3.4)
is essentially based on the idea of incorporating the
constraints (3.8) into the wave equation. Choosing (3.4)
this yields then the formula
\begin{eqnarray} \label{21}
i\partial_0(\beta_0^2 \psi)&=&\left[\beta_0^2H\beta_0^2 -{1 \over m}
\beta_0^2H\beta_i{\cal D}^i\beta_0^2\right](\beta_0^2\psi)
\nonumber \\
&\equiv & {\cal O}_{red}(\beta_0^2\psi)
\end{eqnarray}                                
which should be regarded 
simply as an equation to determine the reduced components $\beta_0^2\psi$.
We do not identify ${\cal O}_{red}$ with
a Hamiltonian, because the operator ${\cal O}_{red}$ is not hermitian. 
A more compact form of the new operator is obtained
by noticing that
\begin{eqnarray} \label{22}
&\beta_0^2&(\beta_{i_1}\beta_{i_2}\ldots \beta{i_{2n+1}})\beta_0^2=0
\nonumber \\
&\beta_0^2&(\beta_{i_1}\beta_{i_2}\ldots \beta{i_{2n}})=
(\beta_{i_1}\beta_{i_2}\ldots \beta{i_{2n}})\beta_0^2=
\beta_0^2(\beta_{i_1}\beta_{i_2}\ldots \beta{i_{2n}})\beta_0^2
\end{eqnarray}
${\cal O}_{red}$ can be then put in the following form
\begin{equation} \label{23}
{\cal O}_{red}=-\beta_0m + \beta_0^2eA_0+{1 \over m}\beta_0\beta_i
\beta_j{\cal D}^i{\cal D}^j
\end{equation}
Since the $\beta_i$ matrices do not have a direct interpretation as operators
in the Hilbert space it would be desireable at this point
to make some contact with spin operators $S_k$. We can do
this by defining $S_k$ through
\begin{equation} \label{24}
S_{ij}=-i\epsilon_{ijk}S_k
\end{equation}
Indeed in our explicit represention, (4.6) implies that
\begin{equation} \label{25}
S_k=\left(\begin{array}{llll}
T_k & 0_{3 \times 3} & 0_{3 \times 3} & \bar{0} \\
0_{3 \times 3} & T_k & 0_{3 \times 3} & \bar{0} \\
0_{3 \times 3} & 0_{3 \times 3} & T_k & \bar{0} \\
\bar{0}^T & \bar{0}^T & \bar{0}^T & 0 \end{array}\right)
\to \left(\begin{array}{ll}
T_k & 0_{3 \times 3} \\
0_{3 \times 3} & T_k \end{array}\right)
\end{equation} 
where $(T_k)_{ij}=i\epsilon_{ikj}$ is the 3-dimensional representation
of $SU(2)$ and $\bar{0}$ is a 3-dimensional zero vector $\bar{0}^T=(0,0,0)$. 
As a result, the correct commutation relation will emerge
\begin{equation} \label{26}
[S_i, S_j]=i\epsilon_{ijk}S_k
\end{equation}
Since 
$\beta_0^2$ projects onto a six dimensional space we have displayed in 
(4.6)
the form of $S_k$ to which it will reduce in this space. Besides the
commutation relation (4.7) it is convenient to have yet another,
preferably anti-commutative relation which would relate the $\beta_i$
to the spin operators $S_k$. One can find such a relation by introducing the
auxiliary matrix 
\begin{equation} \label{27}
\xi =\left(\begin{array}{llll}
1_{3 \times 3} & 0_{3 \times 3} & 0_{3 \times 3} & \bar{0}\\
0_{3 \times 3} & -1_{3 \times 3} & 0_{3 \times 3} & \bar{0} \\
0_{3 \times 3} & 0_{3 \times 3} & -1_{3 \times 3} & \bar{0} \\
\bar{0}^T & \bar{0}^T & \bar{0}^T & 1 \end{array}\right)
\to \left(\begin{array}{ll}
1_{3 \times 3} & 0_{3 \times 3}\\
0_{3 \times 3} & -1_{3 \times 3} \end{array}\right)
\end{equation}
where again the $6 \times 6$ expression of this matrix has been displayed.
By explicit calculation one can check that the identities 
\begin{eqnarray} \label{28}
\{\beta_i, \beta_j\}&=&\xi\{S_i, S_j\}, \,\,\, i \neq j \nonumber \\
\beta_k^2 &=& -{1 \over 2}(1+\xi) +\xi S_k^2
\end{eqnarray}
hold.
Another useful formula involving the spin matrices
follows simply from the fact that we are working with the
adjoint representation $(T_k)_{ij}=
i\epsilon_{ikj}$. It is trivial to see that the $T_k$, and hence the $S_k$,
fullfill an algebra similar to (1.2), namely 
\begin{equation} \label{29}
S_iS_kS_j + S_jS_kS_i =\delta_{ik}S_j + \delta_{jk}S_i
\end{equation}
The Duffin algebra (1.2) is essentially a covariant generalization of (4.10).
Having introduced the spin operators we can rewrite (4.4) into a physically
more suitable form
\begin{equation} \label{30}
{\cal O}_{red}= -\beta_0m +eA_0 +{1 \over 2m}\beta_0(1+\xi)({\cal D}^i
{\cal D}_i) +{1 \over m}\beta_0\xi(S_j{\cal D}^j)^2
+{e \over 2m}\beta_0(1+\xi)(S_kB^k)
\end{equation}
Provided the wave function $\psi$ is an eigenstate to the Hamilton operator
$H$ with an eigenvalue $E \neq 0$, one more step in the reduction
is possible and we can reduce (4.11)
to the first three components of $\beta_0^2\psi$ denoted here by $\chi$.
For simplicity we write the result of this further reduction for 
$A_0=0$
\begin{eqnarray} \label{ex2}
E^2 \chi&=&\biggl [ m^2 -{\cal D}^i{\cal D}_i -{e \over m}(T_kB^k)
\nonumber \\
&-&{1 \over m^2}(T_i{\cal D}_i)^2({\cal D}^j{\cal D}_j)-{1 \over m^2}
(T_kB^k)^4-{e \over m^2}(T_i{\cal D}^i)^2(T_kB^k)\biggl]\chi
\end{eqnarray}
Note that (4.11) is already of second order and as a consequence
eq. (4.12) is of fourth order.

The important conclusion of the reduction process to the physical components
of the wave function is that the troublesome term
$${ie \over 2m}F^{\mu \rho}(\beta_{\rho}\beta_0\beta_{\mu}+\beta_{\rho}
g_{\mu 0})$$      
vanishes, i.e., it reduces to zero. This means that it does not matter
whether one couples the photon via the minimal scheme in the Lagrangian
or as suggested in \cite{ghose3}
in the Hamilton form. The end result of the reduction 
will be the same which implies that the above term does not have any physical
meaning. Actually Kemmer has discussed a similar looking term at hand of 
the second order equation (3.5) \cite{kemmer}. Indeed it is not ruled out
yet that for instance term involving        
${ie \over 2m}F^{ij}(\beta_{j}\beta_k\beta_{i}+\beta_{j}
g_{ik})$ are physically meaningful. This, however, presupposes that (3.5)
itself is meaningful. The next two section are centered around this
problematic.   
\setcounter{equation}{0}
\section{The second order wave equation}

To understand better the
implications of the equation (3.5) it is helpful to have a glimpse at what
happens in the Dirac theory. In the presence of electromagnetic field the Dirac
equation reads
\begin{eqnarray} \label{31}
&\Lambda_{\pm}& \equiv \gamma_{\mu}{\cal D}^{\mu} \pm m \nonumber \\
&\Lambda_-&\Psi =0
\end{eqnarray}
The second order equation is obtained by writing $\Lambda_+\Lambda_-
\Psi=0$ where
\begin{equation} \label{32}
\Lambda_+\Lambda_-={\cal D}^{\alpha}{\cal D}_{\alpha} -m^2 -{e \over 2}
F^{\mu \nu}\sigma_{\mu \nu}
\end{equation}
where as usual $\sigma_{\mu \nu} ={i \over 2}[\gamma_{\mu}, \gamma_{\nu}]$.
Now in the Dirac theory it is simple to pass from solutions of the second order
wave equation $\Psi$ to solutions of the Dirac equation. By virtue of
$$[\Lambda_+, \Lambda_-]=0$$
it follows that
\begin{equation} \label{33}
\Lambda_-(\Phi)=\Lambda_-(\Lambda_+\Psi)=0
\end{equation}
In other words, if $\Psi$ solves $\Lambda_+ \Lambda_- \Psi=0$ then 
$\Phi=\Lambda_+ \Psi$ is a solution of (5.1).
This one-to-one correspondence makes the second order Dirac equation
a meaningful physical equation. Indeed its practical usefulness lies in the
fact that one can easily obtain solutions of the Dirac equation for certain
field configurations by using  (5.2). 
We should therefore try to find out if the successful
Dirac story can formally be
taken over to the case of the second order Kemmer equation.
As a first step in this direction we make an ansatz to factorize the operator
$\Omega_1({\cal D})$ in the form
\begin{equation} \label{34}
d_1({\cal D})\Lambda({\cal D})=\Omega_1({\cal D})
\end{equation}
where $d_1({\cal D})$ is to be found. Such an operator indeed exists and the 
solution is
given by
\begin{equation} \label{35}
d_1({\cal D})={1 \over m}\left[({\cal D}^{\alpha}{\cal D}_{\alpha})
-m^2\right] 
+\beta_{\nu}{\cal D}^{\nu} -{1 \over m}\beta_{\sigma}\beta_{\delta}
{\cal D}^{\delta}{\cal D}^{\sigma}
\end{equation}
This is a straightforward generalization of the operator of 
Takahashi and Umezawa 
\cite{umez} who have found that in the free case
\begin{equation} \label{ex3}
d_1(\partial)\Lambda(\partial)=-(\Box +m^2)
\end{equation}
Although this looks already quite promising the fact that the commutator
\begin{equation} \label{36}
[d_1({\cal D}), \Lambda({\cal D})]={e \over 2m}\beta_{\sigma}\beta_{\rho}
\beta_{\delta}(\partial^{\rho}F^{\delta \sigma}) 
-{ie \over m}\beta_{\sigma}F^{\rho \sigma}{\cal D}_{\rho}
\end{equation}
is non-zero spoils a complete analogy
to the Dirac case. We therefore do not have a clear
prescription for 
how to obtain solutions of the KDP equations out of solutions of 
the second order Kemmer equations (3.5). It could still be, despite (5.7),
that some kind of a map similar to (5.3) exists.
However, the situation is more serious once one
realizes that (3.5) is not unique and belongs to a class of
second order Klein-Gordon type equations. Indeed, we can multiply eq. (3.2)
by some non-zero operator $\tilde{d}({\cal D})$ to arrive at
\begin{equation} \label{ex4}
\tilde{d}({\cal D})\Lambda({\cal D})={\cal D}^{\alpha}{\cal D}_{\alpha}
-m^2 +{\cal G}[A_{\mu}]
\end{equation}               
where ${\cal G}[A_{\mu}]$ is a functional 
of the four-potential which vanishes in the interaction free case.
An explicit example is   
\begin{eqnarray} \label{37}
&d_1&({\cal D}) \to d_1({\cal D})+{ie \over 2m}S_{\rho \sigma}F^{\rho \sigma}
\nonumber \\
&\Omega_1&({\cal D}) \to \Omega_1({\cal D})-{ie \over 2m}
\beta_{\delta}\beta_{\sigma}\beta_{\mu}F^{\sigma \delta}{\cal D}^{\mu}
+{ie \over 2}F^{\mu \nu}S_{\mu \nu}
\end{eqnarray}
Again this shifted operator will not commute with $\Lambda({\cal D})$. We
cannot in analogy to (5.9) redefine $\Lambda_+$ in the Dirac theory without
spoiling the crucial equation (5.3).
A preferable operator to be constructed in the KDP theory 
would be one which satisfies (5.8)
but also $[\tilde{d}({\cal D}), \Lambda({\cal D})]=0$. This is, however, too
strong a requirement. 
Note that the main objective of (5.6) in the free case
is to interpret $m$ as the mass of the particle. Also $d_1(\partial)$
and $\Lambda(\partial)$ commute. If we insist on a higher order wave equation
containing interaction terms we have to
give up the requirement that the product of the two operators is of
Klein-Gordon type (5.8). We will suggest such an equation in the next section.

\setcounter{equation}{0}
\section{The third order wave equation} 

Giving up the requirement in form of eq. (5.8) we are looking for an operator
$d_2({\cal D})$ with the following properties
\begin{eqnarray} \label{38}
d_2({\cal D})\Lambda({\cal D})\psi &=&0 \nonumber \\
{[} d_{2}({\cal D}), \Lambda({\cal D})] &=&0
\end{eqnarray}
We will skip the calculational details and give only the main results here.
Using the algebra (1.2) and on account of $\partial_{\mu}F^{\mu \nu}=0$,
we have
\begin{equation} \label{40}
[\beta_{\mu}, S_{\delta \sigma}]=g_{\mu \delta}\beta_{\sigma}
-g_{\mu \sigma}\beta_{\delta}
\end{equation}
as well as
\begin{equation} \label{42}
\beta_{\mu}\beta_{\delta}\beta_{\sigma}(\partial^{\mu}F^{\delta \sigma})
=-{1 \over 2}\beta_{\sigma}\beta_{\mu}\beta_{\delta}(\partial^{\mu}
F^{\delta \sigma})
\end{equation}
It then follows that 
\begin{eqnarray} \label{39}
&[&{{\cal D}^{\alpha}{\cal D}_{\alpha} \over m}, \Lambda({\cal D})]
=-2{ie \over m}F^{\delta \rho}\beta_{\rho}{\cal D}_{\delta}
\nonumber \\
&[&{ie \over 2m}S_{\rho \sigma}F^{\rho \sigma}, \Lambda({\cal D})]
=-{ie \over m}F^{\rho \sigma}\beta_{\sigma}{\cal D}_{\rho}
-{e \over 2m}\beta_{\sigma}\beta_{\rho}\beta_{\delta}(\partial^{\rho}
F^{\delta \sigma})
\end{eqnarray}
Combining these findings with (5.7) we see immediately that the operator
\begin{equation} \label{43}
d_2({\cal D})\equiv d_1({\cal D})+{ie \over 2m}S_{\delta \sigma}
F^{\delta \sigma}- {{\cal D}^{\alpha}{\cal D}_{\alpha} \over m}
\end{equation}
indeed satisfies the desired commutation relation in (6.1). The equation to be
solved reads now     
\begin{eqnarray} \label{44}
-\Omega_2({\cal D})\psi \equiv -d_2({\cal D})\Lambda({\cal D})\psi &=&
\biggl[m^2 + {ie \over 2m}(\beta_{\sigma}\beta_{\delta}\beta_{\mu}
-\beta_{\sigma}g_{\delta \mu})F^{\mu \sigma}{\cal D}^{\delta}
\nonumber \\
&-& {ie \over 2m}\beta_{\mu}S_{\delta \sigma}{\cal D}^{\mu}F^{\delta \sigma}
+{1 \over m}\beta_{\mu} {\cal D}^{\mu}{\cal D}^{\alpha}{\cal D}_{\alpha}
\biggr]\psi =0
\end{eqnarray}
Essentially 
we have achieved our goal since $\phi \equiv d_2({\cal D})\psi$ is a solution
of (3.2) provided $\psi$ is a solution of (6.6). The last equation
is quite different from
(3.5). It is a third order equation wich upon reduction
becomes a fourth order equation. This is understandable since the reduced
eq. (4.11) 
is already of second order. Note also that the term $F^{\mu \nu}S_{\mu
\nu}$ vanished in (6.6). 

It should be also stressed that (6.6) 
does not have the
limit of the Klein-Gordon equation in the free case. This has consequences for
the interpretation of the interaction terms in (6.6).
In the free case we need
eq. (5.6) 
to interprete $m$ as the mass of the particle. Indeed this is the only
possible interpretation in the free case. Since as we said (6.6) does not have
the Klein-Gordon limit we should not expect that every term in this
equation
will have a simple physical interpretation when we switch on the interaction.
We should regard (6.6) as an equation equivalent to 
(3.2) as far as the solutions
for the wave functions are concerned. 
Due to (3.7) a part of the interaction terms
can, however, be interpreted as spin-field interaction at the order $e^2$. This
is clear after a reduction similar to (4.11) is done.

Although (6.6) is, from the mathematical 
point of view, the counterpart to the second
order Dirac equation (this despite the fact that (6.6) is of third order) it is
clear that its physical implications are quite different. In addition to the
differences discussed above we should also mention that (6.6) will not be
of much practical help to obtain solutions of the KDP equation. This is clear
since it is not a second order Klein-Gordon type equation.
\setcounter{equation}{0}
\section{Conclusions}

We have tried in this letter to shed some light on the non-trivial nature
of the electromagnetic coupling in the KDP equation.
To do so we 
have reduced the KDP equation to the physical components of the wave
function. Along with the reduction process a term which resisted a physical
interpretaion has vanished. We could therefore show that this term does not
have any physical significance. We argued that the second order wave equation
proposed by Kemmer as an analogy to the second order Dirac equation has only a
limited applicablilty (in contrast to the corresponding Dirac case) as
(i) it belongs to a class of possible second order  
equations, (ii) it lacks the important
property which would allow to go back to the solutions of (3.2).
Strictly speaking, the analogy to the second order Dirac equation is then lost.
In view of all these points we do not think that it is really necessary
to give an interpretation of the interaction terms in every such equations.
To obtain a counterpart (at least from the mathematical point of view)
to the second order Dirac equation
we have suggested a third order equation having the desired properties
which allows one to map the solution of the latter equation to solutions
of (3.2),
but lacks a simple physical interpretion since it does not reduce to the
Klein-Gordon equation in the free case.

{\bf Acknowledgments}. I would like to thank A. Bramon, P. Ghose and
M. Lavelle for useful discussions and comments. This work has been 
supported by the Spanish Ministerio de Educacion y Ciencies.

\end{document}